\documentstyle[aps,prb,psfig,multicol]{revtex}

\begin{document}

\def\be{\begin{eqnarray}}
\def\ee{\end{eqnarray}}
\def\nn{\nonumber}
\def\cad{c_A^\dagger}
\def\ca{c_A}
\def\cbd{c_B^\dagger}
\def\cb{c_B}

\title{Spin-Orbit Coupling and Time-Reversal Symmetry in Quantum Gates}

\author{D. Stepanenko and N.E. Bonesteel} 
\address
{Department of
Physics and National High Magnetic Field Laboratory, Florida State
University, Tallahassee, FL 32310} 

\author{D.P. DiVincenzo and G. Burkard}
\address
{IBM Research Division, T.J. Watson Research Center, Yorktown
Heights, NY 10598}

\author{Daniel Loss}
\address
{Department of Physics and Astronomy, University of Basel, Klingelbergstrasse
82, CH-4056 Basel, Switzerland}

\maketitle

\begin{abstract}
We study the effect of spin-orbit coupling on quantum gates produced
by pulsing the exchange interaction between two single electron
quantum dots. Spin-orbit coupling enters as a small spin precession
when electrons tunnel between dots. For adiabatic pulses the resulting
gate is described by a unitary operator acting on the four-dimensional
Hilbert space of two qubits. If the precession axis is fixed,
time-symmetric pulsing constrains the set of possible gates to those
which, when combined with single qubit rotations, can be used in a
simple CNOT construction.  Deviations from time-symmetric pulsing
spoil this construction. The effect of time asymmetry is studied by
numerically integrating the Schr\"odinger equation using parameters
appropriate for GaAs quantum dots. Deviations of the implemented gate
from the desired form are shown to be proportional to dimensionless
measures of both spin-orbit coupling and time asymmetry of the pulse.

\end{abstract}


\begin{multicols}{2}

\section{Introduction}

A promising proposal for building a solid-state quantum computer is
based on the notion of using electron spins trapped in quantum dots as
qubits.\cite{loss} In such a device, two-qubit quantum gates would be
carried out by turning on and off the exchange interaction between
spins on neighboring dots through suitable pulsing of gate voltages.

When performing such a quantum gate, if nonadiabatic errors
\cite{burkard99_1,burkard99_2,hu} can be safely ignored,
\cite{schliemann} both the initial and final states of the two dots
will be in the four-dimensional Hilbert space of two qubits. In the
absence of spin-orbit coupling, and neglecting the dipolar interaction
between spins, the unitary transformation resulting from such a pulsed
exchange gate will necessarily have the form
\be U = \exp -i \lambda {\bf S}_A \cdot {\bf S}_B,
\label{isotropic}
\ee 
where $\lambda$ is a dimensionless measure of the pulse strength.
This simple isotropic form is a consequence of symmetry --- if spin
and space decouple exactly, as they do in the nonrelativistic limit,
then the system is perfectly isotropic in spin space.  Up to an
irrelevant overall phase the gates (\ref{isotropic}) are the most
general unitary operators with this symmetry acting on a two-qubit
Hilbert space.

These isotropic exchange gates are useful for quantum computation.  In
conjunction with single qubit rotations, they can be used in a simple
construction of a controlled-not (CNOT) gate.\cite{loss} It has also
been shown that, even without single qubit rotations, isotropic
exchange gates can be used for universal quantum computing with proper
encoding of logical qubits.\cite{bacon,divincenzo}

When the effects of spin-orbit coupling are included, well-isolated
single electron dots will have a two-fold Kramers degeneracy and so
can still be used as qubits.  However, when carrying out a quantum
gate the total spin will no longer be a good quantum number. As a
result there will inevitably be corrections to the isotropic exchange
gates (\ref{isotropic}).  Motivated by this fact, a number of authors
have considered {\it anisotropic} gates of the form
\be 
U &=& \exp -i \lambda \bigl({\bf S}_A \cdot
{\bf S}_B + {\mbox{\boldmath{$\beta$}}} \cdot({\bf S}_A \times {\bf S}_B)\nn\\
&&~~~~~~~~~~+
\gamma( 
{\bf S}_A \cdot {\bf
S}_B
-
(\hat{\mbox{\boldmath{$\beta$}}}
\cdot{\bf S}_A)
(\hat{\mbox{\boldmath{$\beta$}}}\cdot{\bf S}_B)\bigr),
\label{axial} 
\ee 
and shown that they have several useful properties.  For example, in
Ref.~\onlinecite{burkard01} it was shown that the CNOT construction of
Ref.~\onlinecite{loss} is robust against anisotropic corrections of
the form appearing in (\ref{axial}). It has also been shown that, when
combined with a controllable Zeeman splitting, the gates (\ref{axial})
form a universal set.\cite{wu}

The anisotropic terms which appear in (\ref{axial}) are not the most
general corrections to (\ref{isotropic}) which can occur when carrying
out an exchange gate in the presence of spin-orbit coupling.  It is
therefore important to ask under what conditions these corrections can
be restricted to have this desired form. The key observation
motivating the present work is that, up to an irrelevant overall
phase, the gates (\ref{axial}) are the most general two-qubit quantum
gates which are both axially symmetric, i.e. symmetric under rotations
about an axis parallel to the vector ${\mbox{\boldmath{$\beta$}}}$ in
spin space, and symmetric under time reversal (${\bf S}_\mu
\rightarrow -{\bf S}_\mu$, $\mu = A,B$).  It follows that if these
symmetries can be maintained throughout the gate operation, and
provided nonadiabatic errors can be neglected, the resulting quantum
gate is {\it guaranteed} to have the form (\ref{axial}).  Of course
symmetry alone cannot determine the values of $\lambda$,
${\mbox{\boldmath{$\beta$}}}$ and $\gamma$.  However, in practice we
envision these parameters will be determined through experimental
calibration rather than microscopic calculation.  Therefore we
emphasize symmetry as a useful guiding principle.

In this paper we study the effect of spin-orbit coupling on
exchange-based quantum gates.  For concreteness we consider a system
of two single-electron quantum dots in GaAs. The contribution of
spin-orbit coupling to the exchange interaction between localized
spins in GaAs has been studied by Kavokin\cite{kavokin} within the
Heitler-London approximation, and by Gor'kov and Krotkov\cite{gorkov}
who derived the exact asymptotic exchange interaction between
hydrogen-like bound states at large separation.

Here we follow Ref.~\onlinecite{burkard99_1} and work within the
Hund-Mulliken approximation, keeping one orbital per dot, and allowing
double occupancy.  In this approximation, the effect of spin-orbit
coupling is to induce a small spin precession whenever an electron
tunnels from one dot to another.  The Hamiltonian governing the
two-dot system is therefore axially symmetric in spin space with the
symmetry axis being the precession axis of the spin.  If the direction
of the precession axis does not change while the gate is being pulsed,
then the resulting quantum gate will also be axially symmetric.

An additional useful symmetry principle, first suggested in
Ref.~\onlinecite{bonesteel01}, is that any time-dependent Hamiltonian
$H_P(t)$ which is time-reversal symmetric at all times $t$, and which
is then pulsed in a time-symmetric way ($H_P(t) = H_P(-t)$) will lead
to a gate which can be described in terms of an effective
time-independent Hamiltonian $H$ which is also time-reversal
symmetric.  Here we give a proof of this result.

Taken together these two results imply that, within the Hund-Mulliken
approximation, if the spin-orbit precession axis is fixed and
nonadiabatic errors can be ignored, the unitary transformation
produced by pulsing the exchange interaction between two quantum dots
will necessarily have the desired form (\ref{axial}) provided the gate
is pulsed in a time-symmetric way.

This paper is organized as follows.  In Sec.~II we derive the
Hund-Mulliken Hamiltonian for a double quantum dot system in the
presence of spin-orbit coupling.  In Sec.~III we develop an effective
spin Hamiltonian description which can be applied to pulsing our
double dot system, and we review the robust CNOT construction of
Ref.~\onlinecite{burkard01}.  The implications of time-symmetric
pulsing are then studied in Sec.~IV, and in Sec.~V we present
numerical results showing the effect of small time asymmetry of the
pulse.  Finally, in Sec.~VI we summarize the results of the paper.

\section{Hund-Mulliken Hamiltonian}

We consider a system of two laterally confined quantum dots with one
electron in each dot.  For concreteness we assume the dots are formed
in a two-dimensional electron gas (2DEG) realized in a GaAs
heterostructure.

The system is modeled by the Hamiltonian
\be H = T + C + H_{SO}. 
\label{hamiltonian}
\ee
Here $T+C$ is the Hamiltonian studied in
Ref.~\onlinecite{burkard99_1}, where $T = \sum_i h_i$ with
\be h_i = \frac{1}{2m} \left({\bf p}_i -
\frac{e}{c}{\bf A}({\bf r}_i)\right)^2 + V({\bf r}_i),  \ee
and $C = e^2/\epsilon|{\bf r}_1 - {\bf r}_2|$ is the Coulomb repulsion
between electrons.  We take the 2DEG the dots are formed in to lie in
the $xy$ plane, and for GaAs we take $m = 0.067 m_e$ and $\epsilon =
13.1$.  For completeness we include a vector potential ${\bf A} =
(-y,x,0)B/2$ which couples the orbital motion of the electrons to a
uniform magnetic field ${\bf B} = B \hat{\bf z}$.  We will see in
Sec.~III that this orbital coupling does not affect any of our
arguments based on time-reversal symmetry, while a nonzero Zeeman
coupling does.

As in Ref.~\onlinecite{burkard99_1} lateral confinement of the dots is
modeled by the double-well potential,
\be
V(x,y) = \frac{m\omega_0^2}{2} \left(\frac{1}{4a^2}(x^2 - a^2)^2 +
y^2 \right).
\ee
This potential describes two quantum dots sitting at the points $(x,y)
= (\pm a, 0)$. In the limit of large separation the dots decouple into
two harmonic wells with frequency $\omega_0$.

Spin-orbit coupling enters the Hamiltonian through the term
\be
H_{SO} = \sum_{i=1,2}{\bf \Omega}({\bf k}_i) \cdot {\bf S}_i,
\ee
where $\hbar{\bf k} = {\bf p} - \frac{e}{c} {\bf A}$. Time-reversal
symmetry requires that ${\bf \Omega}({\bf k})$ is an odd function of
${\bf k}$, ${\bf \Omega}({\bf k}) = - {\bf \Omega}(-{\bf k})$.  Thus
${\bf \Omega}$ is nonzero only in the absence of inversion symmetry.

For definiteness, we take the 2DEG in which the dots are formed to lie
in the plane perpendicular to the [001] structural direction, which
then points along the $z$-axis.  However, we allow the $x$-axis, which
is parallel to the displacement vector of the two dots, to have any
orientation with the respect to the [100] and [010] structural axes.
To describe the dependence of ${\bf \Omega}$ on ${\bf k}$ it is then
convenient to introduce unit vectors ${\hat {\bf e}}_{[110]}$ and
${\hat {\bf e}}_{[{\overline 1}10]}$ which point in the $[110]$ and
$[\overline 110]$ structural directions, respectively, and define
$k_{[110]} = {\bf k} \cdot \hat {\bf e}_{[110]}$ and $k_{[\overline
110]} = {\bf k} \cdot \hat {\bf e}_{[\overline 110]}$.  We then have,
following Kavokin,\cite{kavokin}
\begin{eqnarray}
{\bf \Omega}({\bf k}) \simeq 
(f_D - f_R) k_{[110]} \hat {\bf e}_{[\overline 1 1 0]}
+(f_D + f_R)  k_{[\overline 110]} 
\hat {\bf e}_{[110]}.
\label{omega}
\end{eqnarray}
Here $f_D$ is the Dresselhaus contribution\cite{dresselhaus,dyakonov}
due to the bulk inversion asymmetry of the zinc-blende crystal
structure of GaAs, and $f_R$ is the Rashba contribution\cite{rashba}
due to the inversion asymmetry of the quantum well used to form the
2DEG.  These quantities depend on details of the 2DEG confining
potential and so will vary from system to system.

It was pointed out in Ref.~\onlinecite{schliemann03} that $H_{SO}$ has
a special symmetry when $f_D = \pm f_R$.  This can be seen directly
from (\ref{omega}).  When $f_D = f_R$ ($f_D = - f_R$) the direction of
${\bf \Omega}$ is independent of ${\bf k}$ and is fixed to be parallel
to $\hat {\bf e}_{[110]}$ ($\hat {\bf e}_{[\overline 110]}$). The full
Hamiltonian (\ref{hamiltonian}) is then invariant under rotations in
spin space about this axis.  We will see below that this special case
has a number of attractive features.

In the limit of decoupled dots, and ignoring spin-orbit coupling, the
single electron ground states will be the Fock-Darwin ground states
centered at $(x,y) = (\pm a,0)$,
\be \phi_{\pm a}(x,y) = \sqrt{\frac{m\omega}{\pi\hbar}} e^{ -m\omega
\left((x \mp a)^2 + y^2\right)/2\hbar} e^{ \pm 
i a y/2 l_B^2}. \ee
Here $\omega^2 = \omega_0^2 + \omega_{L}^2$ is the frequency of the
magnetically squeezed oscillator where $\omega_{L} = eB/2mc$ is the
Larmor frequency and $l_B = \sqrt{\hbar c/eB}$ is the magnetic length.
In zero magnetic field, the size of these wave functions is set by the
effective ``Bohr radius'' $a_B = \sqrt{\hbar/m\omega_0}$.

The Fock-Darwin states can be orthogonalized to obtain the Wannier
states
\be
\Phi_A &=& \frac{1}{\sqrt{1-2Sg -g^2}} (\phi_a - g \phi_{-a}),\\
\Phi_B &=& \frac{1}{\sqrt{1-2Sg -g^2}} (\phi_{-a} - g \phi_{a}),
\ee
where $S = \langle \phi_{-a} | \phi_{a} \rangle$ and $g =
(1-\sqrt{1-S^2})/S$.  We can then introduce second quantized operators
${c^\dagger_A}_\alpha$ (${c_A}_\alpha$) and ${c^\dagger_B}_\alpha$
(${c_B}_\alpha$) which create (annihilate) electrons in the states
$\Phi_A$ and $\Phi_B$ with spin $\alpha = \uparrow,\downarrow$.

In the Hund-Mulliken approximation we keep one orbital per dot and
allow for double occupancy.  This amounts to restricting the full
Hilbert space of the problem to the six-dimensional Hilbert space
spanned by the states
\be |S_1 \rangle &=& \frac{1}{\sqrt{2}} (c^\dagger_{A\uparrow}
c^\dagger_{B\downarrow} -c^\dagger_{A\downarrow}
c^\dagger_{B\uparrow}) |0\rangle,\\ |S_2 \rangle &=& \frac{1}{\sqrt{2}}
(c^\dagger_{A\uparrow} c^\dagger_{A\downarrow}
+c^\dagger_{B\downarrow} c^\dagger_{B\uparrow}) |0\rangle,\\
|S_3\rangle &=& \frac{1}{\sqrt{2}} (c^\dagger_{A\uparrow}
c^\dagger_{A\downarrow} -c^\dagger_{B\downarrow}
c^\dagger_{B\uparrow})|0\rangle,\\ 
|T_-\rangle &=&
c^\dagger_{A\downarrow}c^\dagger_{B\downarrow} |0\rangle,\\ |T_0\rangle
&=& \frac{1}{\sqrt{2}} (c^\dagger_{A\uparrow} c^\dagger_{B\downarrow}
+c^\dagger_{A\downarrow} c^\dagger_{B\uparrow}) |0\rangle,\\
|T_+\rangle &=& c^\dagger_{A\uparrow} c^\dagger_{B\uparrow} |0
\rangle.  \ee

In terms of second quantized operators, the Hund-Mulliken Hamiltonian
acting in this space, up to an irrelevant overall additive
constant, can be written
\be H_{HM} &=& \sum_{\alpha,\beta = \uparrow,\downarrow}
-\left({c^\dagger_A}_\alpha (t_H \delta_{\alpha\beta} + i {\bf P} \cdot
{\mbox{\boldmath{$\sigma$}}}_{\alpha\beta}){c_B}_\beta+H.c.\right)\nn\\
&&+V\left({\bf S}_A \cdot {\bf S}_B+3/4\right)\nn\\ &&+ U_H
(n_{A\uparrow} n_{A\downarrow} + n_{B\uparrow} n_{B\downarrow}).  \ee
\noindent
Here
\be
{\bf S}_{\mu} = \frac{1}{2} \sum_{\alpha,\beta = \uparrow,\downarrow}
{c^\dagger_\mu}_\alpha {\mbox{\boldmath$\sigma$}}_{\alpha\beta} {c_\mu}_\beta
\ee
is the spin operator on site $\mu = A,B$, 
\be
V &=& \langle S_1 | C | S_1 \rangle - \langle T | C| T \rangle
\ee
is the ferromagnetic direct exchange,
\be
U_H &=& \langle S_2 | C| S_2 \rangle - \langle S_1 | C | S_1 \rangle
\ee
is the Coulomb energy cost of doubly occupying a dot,
and 
\be
t_H &=& \langle \Phi_A | h | \Phi_B \rangle
\ee
is the interdot tunneling amplitude.  

The only contribution from spin-orbit coupling is the matrix element
\be i{\bf P} = \langle \Phi_A | {\bf \Omega}({\bf k}) | \Phi_B
\rangle  = \langle \Phi_A | (p_x - \frac{e}{c}A_x)/\hbar | \Phi_B
\rangle
\mbox{\boldmath{$\eta$}}, \ee
where 
\be
\mbox{\boldmath{$\eta$}} &=& (f_D - f_R) \cos \theta \hat {\bf e}_{[\overline 110]}+ (f_D + f_R ) \sin \theta \hat {\bf e}_{[110]}.
\ee
Here $\theta$ is the angle the $x$-axis makes with the [110]
structural direction.  This term introduces a small spin precession
about an axis parallel to ${\bf P}$ through an angle $\phi = 2
\arctan(P/t_H)$ when an electron tunnels between dots.

It is convenient to express the spin-orbit matrix element as ${\bf
P}=s{\bf l}_{SO}$ where
\be s=\frac{\sqrt{(f_D-f_R)^2 \cos^2 \theta + (f_D+f_R)^2 \sin^2
\theta }}{a_B~\hbar\omega_{0}} 
\label{sdef}
\ee
is a dimensionless measure of the strength of spin-orbit coupling.  As
stated above, $f_D$ and $f_R$ depend on details of the potential
confining the electron to the 2DEG. Thus $\theta$, $f_D$ and $f_R$ are
all parameters that in principle can be engineered to control the
value of $s$.  For example, if $\theta = 0$ then $s = |f_D - f_R|/(a_B
\hbar\omega_0)$.  Thus, for this orientation of the dots, if it is
possible to design a system in which $f_D = f_R$, $s$ can be made to
vanish. Even if such perfect cancellation cannot be achieved,
minimizing the difference $f_D - f_R$ will reduce $s$.

\begin{figure}[h]
\centerline{\psfig{figure=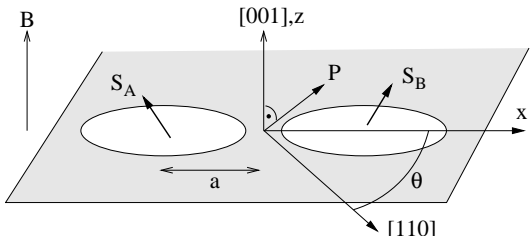,height=1.2in,angle=0}}\vskip .15in
\caption{Sketch of the GaAs double quantum dot system considered in
this paper.  There is one electron per dot, and the dot separation is
$2a$.  The dots are taken to lie in the plane perpendicular to both
the [001] axis and an applied magnetic field $B$.  The displacement of
the dots makes an angle $\theta$ with the [110] axis.  Due to
spin-orbit coupling electron spins precess about an axis parallel to
${\bf P}$ when tunneling between dots.}
\label{system}
\end{figure}

In what follows we leave $s$ as a free parameter.  We estimate that
for GaAs quantum dots $s < 0.1$ for typical parameters.\cite{kavokin}
The remaining contribution to the matrix element ${\bf P}$ is then
\be {\bf l}_{SO}=\frac{\hbar\omega_0}{2} \frac{1-g^2}{1-2Sg+g^2} \frac{d}{b}
e^{-d^2 b(2-1/b^2)} \hat{\mbox{\boldmath{$\eta$}}}, \ee
where $d = a/a_B$ is a dimensionless measure of the distance between
dots, $b = \sqrt{1+\omega_L^2/\omega_0^2}$, and
$\hat{\mbox{\boldmath{$\eta$}}} = \mbox{\boldmath{$\eta$}}/\eta$.  The
geometry of our model system is shown schematically in
Fig.~\ref{system}.

In what follows we envision pulsing quantum gates by varying the
distance $d$ between dots as a function of time.  In doing this, we
will assume that throughout the pulse the values of $f_D$ and $f_R$ do
not change.  If this is the case $s$ will be constant and all of the
time dependence of ${\bf P}$ will be due to ${\bf l}_{SO}$.  In
addition, the direction of the vector ${\bf P}$ will not change as a
function of time.  The Hamiltonian $H_{HM}$ will therefore be
invariant under rotations in spin space about a single fixed axis
parallel to ${\bf P}$ throughout the pulse.  We will refer to such a
pulse as having axial symmetry.
 
\
\vskip .2in

It is important to note that this axial symmetry is approximate.  In
general $f_D$ and $f_R$ will depend on time as the gate is pulsed,
though in principle the system can be engineered to minimize this
effect.  Also, for general $f_D$ and $f_R$ the appearance of only one
vector in spin space is a consequence of restricting the Hilbert space
to one orbital per dot.  If more orbitals are included then more
spin-orbit matrix elements will appear in the Hamiltonian,
corresponding typically to different spin-precession axes, thus
breaking the axial symmetry.  However, as shown above, if $f_D = \pm
f_R$ then the full Hamiltonian (\ref{hamiltonian}) is axially
symmetric -- thus for this special case all spin precession axes will
be parallel and axial symmetry will not just be an artifact of the
Hund-Mulliken approximation.  In Sec.~V we discuss the effect
deviations from axial symmetry will have on our results.

Given an axially symmetric pulse, it is convenient to take the
$z$-axis in spin space to be parallel to ${\bf P}$.  For this choice,
the states $|T_+\rangle$ and $|T_-\rangle$ decouple, each having
energy $V$.

Another useful symmetry of $H_{HM}$ is invariance under $c_{A\alpha}
\rightarrow c_{B,-\alpha}$ and $c_{B\alpha} \rightarrow
c_{A,-\alpha}$.  This transformation changes the sign of the states
$|S_1\rangle,\ |S_2\rangle$ and $|T_0\rangle$, while leaving
$|S_3\rangle$ invariant.  It follows that the state $|S_3\rangle$ also
decouples with energy $U_H$.  The matrix representation of $H_{HM}$ in
the remaining nontrivial $|T_0\rangle$, $|S_1\rangle$, $| S_2\rangle$
basis is then
\be H_{HM} = \left(
\begin{array}{ccc}
V & 0 & -2iP \\
0 & 0 & -2t_H \\
2iP & -2t_H & U_H
\end{array}
\right).
\label{hmatrix}
\ee 

\section{Effective Spin Hamiltonian}

We now consider pulsing the Hamiltonian $H_{HM}$ by varying the
distance between the dots, the barrier height, or some combination of
the two, in such a way that the two electron spins interact for a
finite period of time, but are well separated at the beginning and end
of the pulse.  We assume the initial state of the system is in the
four-dimensional Hilbert space describing two qubits, i.e. the space
spanned by the singly occupied states $|S_1\rangle, |T_0\rangle,
|T_-\rangle$ and $|T_+\rangle$.  As the pulse is carried out, the
eigenstates of $H_{HM}$ at any given instant in time can be grouped
into four low-energy states separated by a gap of order $U_H$ from two
high-energy states.  If the pulse is sufficiently adiabatic on a time
scale set by $\sim \hbar/U_H$, the amplitude for nonadiabatic
transitions which would leave the system in the excited state
$|S_2\rangle$ at the end of the pulse can be made negligibly small.
\cite{schliemann} If this condition holds, the final state of the
system can also be assumed to be in the four-dimensional Hilbert space
of two qubits.  We will see that this condition is easily achieved in
Sec.~V.

One way to theoretically study the effect of such a pulse would be to
first reduce $H_{HM}$ to an effective anisotropic spin Hamiltonian
acting on the four-dimensional low-energy Hilbert space and then
consider pulsing this effective model.\cite{bonesteel01} The problem
with this approach is that any such effective spin Hamiltonian will
only be valid if the pulse is adiabatic, not only on the time scale
$\hbar/U_H$, but also on the much longer time scale set by the inverse
of the small energy splittings within the low-energy space due to the
spin-orbit induced anisotropic terms.  However, it is precisely the
nonadiabatic transitions induced by these terms which give rise to the
quantum gate corrections we would like to compute.

Although we may not be able to define an instantaneous effective spin
Hamiltonian during the pulse, we can define one which describes the
net effect of a full pulse.  This definition amounts to parameterizing
the quantum gate produced by the pulse as
\be U = e^{ - i \tau H}, 
\ee
where $U$ acts on the four-dimensional Hilbert space of the initial
and final spin states.  $H$ is then an effective spin Hamiltonian,
i.e. it can be expressed entirely in terms of the spin operators ${\bf
S}_A$ and ${\bf S}_B$, and $\tau$ is a measure of the pulse duration.
Note the definition of $\tau$ is arbitrary because it is the product
$\tau H$ which determines $U$.  Here, and in the remainder of this
paper, we work in units in which $\hbar = 1$.

If we assume exact axial symmetry throughout the pulse, the effective
spin Hamiltonian must be invariant under rotations about the $z$-axis
in spin space and must also leave the states $|T_+\rangle$ and
$|T_-\rangle$ degenerate.  The most general such spin Hamiltonian, up
to an irrelevant additive term proportional to the identity operator,
is
\begin{eqnarray}
\tau { H}
(\lambda;\alpha,\beta,\gamma) &=& \lambda\bigl({\bf S}_A \cdot {\bf S}_B + 
\frac{\alpha}{2}({S_A}_z- {S_B}_z)\nonumber\\
&&~~~~~+{
{\beta}} ( {S_A}_x {S_B}_y - {S_A}_y {S_B}_x)\nonumber\\ 
&&~~~~~+{\gamma} ( {S_A}_x {S_B}_x + {S_A}_y {S_B}_y)\bigr),
\label{hlabg}
\end{eqnarray}
and we denote the corresponding quantum gate as
\be
U(\lambda;\alpha,\beta,\gamma) = e^{-i\tau H
(\lambda;\alpha,\beta,\gamma)}.
\label{axialnt}
\ee
When $\alpha = 0$, this is precisely the gate (\ref{axial}) for
$\mbox{\boldmath{$\beta$}}\parallel \hat {\bf z}$.

The CNOT construction originally proposed in Ref.~\onlinecite{loss} is
based on the sequence of gates
\be
U_g = U(\pi/2; 0,0,0) e^{i\pi {S_A}_z} U(\pi/2; 0,0,0),
\label{cnot}
\ee
where $U(\pi/2; 0,0,0) = \exp-i[(\pi/2) {\bf S}_A \cdot {\bf S}_B]$ is
a square-root of swap gate.  The CNOT gate is then
\be
U_{CNOT} =
e^{i(\pi/2){S_A}_z} e^{i(\pi/2){S_B}_z} U_g.
\ee
Remarkably, it was shown in Ref.~\onlinecite{burkard01} that if
$\lambda = \pi/2$ and $\alpha = 0$ this construction is robust against
the $\beta$ and $\gamma$ corrections, i.e. the gate
\be
U_g = U(\pi/2; 0,\beta,\gamma) e^{i\pi {S_A}_z} U(\pi/2; 0,\beta,\gamma)
\ee
is independent of $\beta$ and $\gamma$.

For completeness, we briefly review the arguments of
Ref.~\onlinecite{burkard01}.  Due to axial symmetry, the action of the
gate $U(\lambda;\alpha,\beta,\gamma)$ on the states $|T_+\rangle$ and
$|T_-\rangle$ is trivial and independent of $\alpha, \beta$ and
$\gamma$,
\be
U(\lambda;\alpha,\beta,\gamma) |T_\pm\rangle =  e^{-i\lambda/4} |T_\pm\rangle.
\ee
We can then introduce a pseudospin description of the remaining space,
where $|S_1\rangle$ is pseudospin down and $|T_0\rangle$ is pseudospin
up. The action of the gate $U(\lambda;\alpha,\beta,\gamma)$ on this
pseudospin space is a simple rotation,
\be U(\lambda;\alpha,\beta,\gamma) \Rightarrow e^{i\lambda/4} 
e^{-i {{\bf b}}\cdot {\mbox{\boldmath$\tau$}}/2}, \ee
where ${\bf b} = \lambda (\alpha,\beta,\gamma+1)$ and the components
of ${\mbox{\boldmath{$\tau$}}} = (\tau_x,\tau_y,\tau_z)$ are pseudospin
Pauli matrices. At the same time, the action of the single qubit
rotation entering $U_g$ is
\be
e^{i\pi {S_A}_z} \Rightarrow i \tau_x.
\ee

Thus to show that the CNOT construction is independent of $\beta$ and
$\gamma$ if $\alpha = 0$ we need only show that the product
\be e^{ -i {\bf b}\cdot {\mbox{\boldmath$\tau$}}/2} \tau_x 
e^{-i {\bf b}\cdot \mbox{\boldmath$\tau$}/2} \ee
is independent of $\beta$ and $\gamma$ if $\alpha = 0$. This condition
has a simple geometric interpretation. It is the requirement that a
rotation about an axis parallel to ${\bf b}$, followed by a 180$^o$
rotation about the $x$-axis, and then a repeat of the initial rotation
must be equivalent to a simple 180$^o$ rotation about the $x$-axis.
This will trivially be the case if the vector ${\bf b} =
\lambda(\alpha,\beta,\gamma+1)$ lies in the $yz$ plane.  Thus, if
$\alpha =0$, this condition is satisfied and the CNOT construction is
exact.  Conversely, if $\alpha \ne 0$ the construction is spoiled.

\section{Time-Reversal Symmetry}

In this section we prove the following general result. Any
time-dependent Hamiltonian $H_P(t)$ which is time-reversal symmetric
for all $t$, and for which the time dependence is itself symmetric,
i.e. $H_P(t_0-t) = H_P(t_0+t)$ for all $t$, will generate a unitary
evolution operator $U = \exp -i \tau H$ where $H$ is a
time-independent effective Hamiltonian which is also time-reversal
symmetric.  We then show that this theorem implies that the parameter
$\alpha$, which spoils the CNOT construction described in Sec.~III, is
equal to zero for time-symmetric pulsing.

The time-reversal operation for any quantum system can be represented
by an antiunitary operator $\Theta$.\cite{gottfried} An orthonormal
basis $\{|M_i\rangle\}$ for the Hilbert space of this system is then
said to be a time-symmetric basis if
\be \Theta | M_i\rangle = |M_i\rangle 
\ee 
for all $i$.

For any Hamiltonian $H$ acting on a state $|M_i\rangle$ in this basis
we can write
\be H |M_i\rangle = \sum_j \langle M_j | H | M_i\rangle
|M_j \rangle. 
\label{hm1}
\ee 
Under time reversal $H$ is transformed into $\Theta H \Theta^{-1}$.
Using the invariance of the $\{|M_i\rangle\}$ basis and the antiunitarity
of $\Theta$ we can then also write
\be \Theta H \Theta^{-1} |M_i\rangle &=&
\Theta H |M_i\rangle \\ &=& \Theta \sum_j \langle M_j | H | M_i\rangle
|M_j \rangle \\ &=& \sum_j \langle M_j | H | M_i\rangle^* |M_j\rangle.
\label{hm2}
\ee 
Comparing (\ref{hm1}) and (\ref{hm2}) leads to the conclusion that if
$H$ is time-reversal symmetric, i.e. $H = \Theta H \Theta^{-1}$, then
the Hamiltonian matrix is purely real in the $\{|M_i\rangle\}$ basis,
while if $H$ is antisymmetric under $\Theta$, i.e. $H = - \Theta H
\Theta^{-1}$, then the Hamiltonian matrix is purely imaginary.

Since $H$ is real in the $\{|M_i\rangle\}$ basis if and only if $H$ is
time-reversal symmetric it follows that the unitary operator $U = \exp
- i\tau H$ is self-transpose, i.e. $U = U^{T}$, if and only if $H$ is
time-reversal symmetric.

Now consider a time-dependent pulse described by the Hamiltonian
$H_P(t)$.  We assume that $H_P(t)$ is time-reversal symmetric at all
times, i.e. $H_P(t) = \Theta H_P(t) \Theta^{-1}$ for all $t$.  The
corresponding unitary evolution operator $U$ which evolves the system
from time $t_I$ to $t_F$ can be written
\begin{eqnarray}
U = \lim_{N\rightarrow\infty}U(t_N) U(t_{N-1}) \cdots U(t_2) U(t_1)
\end{eqnarray}
where 
\begin{eqnarray}
U(t_i) = e^{- i \Delta t H_P(t_i)},
\end{eqnarray}
with $\Delta t = (t_F- t_I)/N$ and $t_1 \equiv t_I$ and $t_N \equiv
t_F$.

Since $H_P(t_i)$ is time-reversal symmetric the above arguments imply
$U^{T}(t_i) = U(t_i)$ when $U(t_i)$ is expressed in the
time-symmetric basis $\{|M_i\rangle\}$. Thus, in this basis, we have
\begin{eqnarray}
U^{T} &=& \lim_{N\rightarrow\infty}
\left( U(t_N) U(t_{N-1}) \cdots U(t_{2}) U(t_{1})
\right)^{T} \\ &=& 
\lim_{N\rightarrow\infty}
U^{T}(t_1) U^{T}(t_{2}) \cdots U^{T}(t_{N-1})
U^{T}(t_{N}) \\ &=& 
\lim_{N\rightarrow\infty}
U(t_1) U(t_{2}) \cdots U(t_{N-1}) U(t_{N}).
\label{transpose}
\end{eqnarray}

For a time-symmetric pulse $H_P(t_i) = H_P(t_{N+1-i})$ and so $U(t_i)
= U(t_{N+1-i})$.  This allows us to reverse the order of the operators
in (\ref{transpose}) which then implies
\begin{eqnarray}
U^{T} = U.
\end{eqnarray}
Thus if we write $U$ in terms of an effective Hamiltonian, 
\be U = e^{
-i \tau {H} },
\ee 
the matrix elements of $H$ must be real in the time-symmetric basis.
$H$ must therefore be time-reversal symmetric, i.e. $H = \Theta H
\Theta^{-1}$.

To apply this theorem to the present problem we take the time-reversal
operator for our two-electron system  to be
\be
\Theta = e^{i\pi {S_A}_y} e^{i\pi {S_B}_y}  K.
\ee
Here the antiunitary operator $K$ is defined so that when acting on a
given state it takes the complex conjugate of the amplitudes of that
state when expressed in the Hund-Mulliken basis defined in Sec.~II.
Note that this basis is constructed using the Fock-Darwin states, and
if a magnetic field is present these states will be necessarily
complex valued when expressed in the position basis.  As defined here,
the antiunitary operator $K$ only takes the complex conjugates of the
amplitudes in the Hund-Mulliken basis, {\it it does not take the
complex conjugate of the Fock-Darwin states themselves}. Thus, if a
magnetic field is present, $\Theta$ should be viewed as an {\it
effective} time-reversal symmetry operator.  This is a technical point
which does not affect any of our conclusions (provided the Zeeman
coupling can be ignored --- see below).  The key property that we will
need in what follows is that spin changes sign under time reversal,
and it is readily verified that for our definition of $\Theta$,
\be
\Theta {\bf S}_\mu \Theta^{-1} = - {\bf S}_\mu
\ee
for $\mu=A,B$ even in the presence of a magnetic field.

Under $\Theta$, the Hund-Mulliken basis states transform as follows,
\be
\Theta |S_i\rangle &=& \phantom{-}|S_i\rangle\ \ \ \ \ {\rm for\ }i=1,2,3,\\
\Theta |T_0\rangle &=& -|T_0\rangle, \\
\Theta |T_+\rangle &=& \phantom{-}|T_-\rangle, \\
\Theta |T_-\rangle &=& \phantom{-}|T_+\rangle.
\ee
The states $|S_i\rangle$ therefore form a time-symmetric basis for the
singlet states.  A time-symmetric basis for the triplet states is
given by
\be 
|\tilde T_0 \rangle &=& i|T_0\rangle, \\
|\tilde T_a \rangle &=& \frac{1}{\sqrt{2}} (|T_+\rangle + |T_-\rangle), \\
|\tilde T_b \rangle &=& \frac{i}{\sqrt{2}}(|T_-\rangle - |T_+\rangle),
\ee  
all of which are eigenstates of $\Theta$ with eigenvalue $+1$.

The matrix representation of $H_{HM}$ in the time-reversal invariant
$|\tilde T_0\rangle$, $| S_1\rangle$, $| S_2\rangle$ basis is
\be H_{HM} = \left(
\begin{array}{ccc}
V & 0 & -2P \\
0 & 0 & -2t_H \\
-2P & -2t_H & U_H
\end{array}
\right),
\label{hmatrix2}
\ee 
which is real, reflecting the effective time-reversal symmetry of
$H_{HM}$.  Note that this would not be the case if $H_{HM}$ included
the Zeeman coupling of electron spins to an external magnetic field.
While for typical field strengths the Zeeman coupling is
small,\cite{burkard99_1} for some parameters it can be comparable to
the spin-orbit corrections considered here.  If this is the case our
conclusions following from effective time-reversal symmetry will no
longer be valid. Of course in zero magnetic field exact time-reversal
symmetry is guaranteed.

We now consider pulsing a time dependent $H_{HM}(t)$ adiabatically so
that, according to the arguments of Sec.~III, the resulting gate can
be parametrized by an effective spin Hamiltonian $H$.  Since at all
times $t$ the Hund-Mulliken Hamiltonian is time-reversal symmetric, if
the pulse itself is time symmetric, i.e. $H_{HM}(t) = H_{HM}(-t)$
where we take the center of the pulse to be at $t=0$, then the above
theorem implies that the effective spin Hamiltonian $H$ will also be
time-reversal symmetric. Thus $H = \Theta H \Theta^{-1}$, and since
$\Theta {\bf S}_\mu \Theta^{-1} = - {\bf S}_\mu$ this implies $H$ must
be quadratic in the spin operators, and so $\alpha = 0$.  The
resulting gate will therefore have the desired form (\ref{axial}).

For completeness we also consider here the case of time-antisymmetric
pulsing.  If $H_P(t) = - H_P(-t)$ then 

\end{multicols}

\begin{figure}[h]
\centerline{\psfig{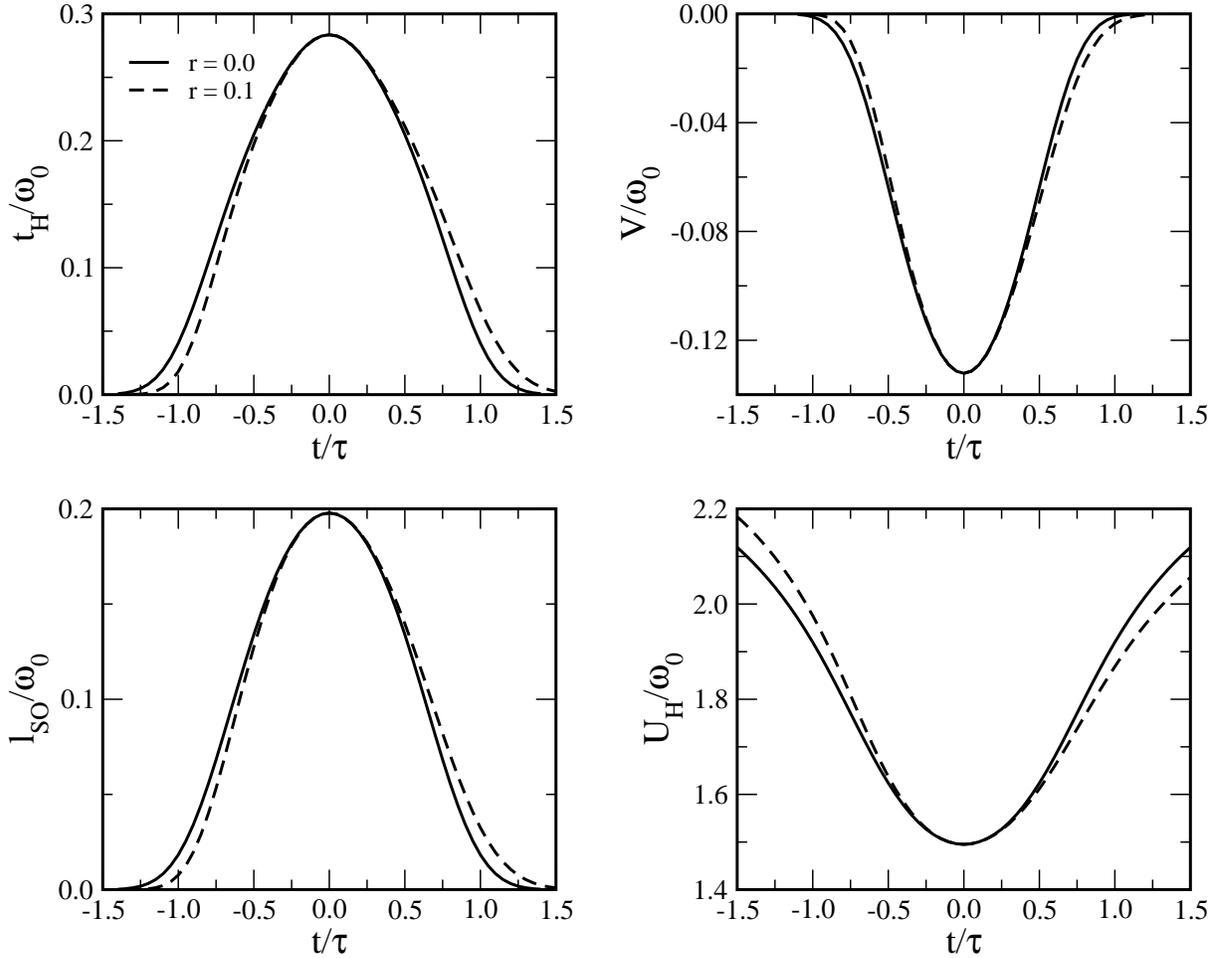}}\vskip .15in
\caption{Time dependence of matrix elements appearing in the
Hund-Mulliken description of a double quantum dot when the
displacement of the dots is varied according to (\ref{doft}) with $d_0
= 1$.  Results are for GaAs parameters in zero magnetic field with
$\hbar\omega_0 = 3$ meV and are plotted vs. the dimensionless quantity
$t/\tau$ for two values of the time-asymmetry parameter, $r = 0$
(solid line) and $r=0.1$ (dashed line).}
\label{dpulse}
\end{figure}

\begin{multicols}{2}

\begin{eqnarray}
U(t) = e^{- i \Delta t H_P(t) } = e^{i \Delta t H_P(-t)}  = U(-t)^{-1},
\end{eqnarray}
and the resulting quantum gate is
\begin{eqnarray}
U &=& \lim_{N\rightarrow\infty}U(t_1) U(t_2) \cdots U(t_{N/2})\nn\\
&&~~~~~~~~~\times U(t_{N/2})^{-1} \cdots U(t_2)^{-1} U(t_1)^{-1} = 1.
\end{eqnarray}
The net effect of any time-antisymmetric pulse is thus simply the
identity transformation.

\section{Model Calculations}

We have seen from symmetry arguments that time-symmetric pulsing of an
axially symmetric Hamiltonian, such as $H_{HM}$ when $f_D$ and $f_R$
are constant, which is itself time-reversal symmetric at all times,
will automatically produce a gate of the form (\ref{axial}), provided
the pulse is adiabatic so that the initial and final states of the
system are in the four-dimensional Hilbert space of two qubits.  It is
natural to then ask what the effect of the inevitable deviations from
time-symmetric pulsing will be on the resulting gate.  To investigate
this we have performed some simple numerical simulations of coupled
quantum dots.

In our calculations, we imagine pulsing the dots by varying the
dimensionless distance $d$ between them according to
\be
d(t) = d_0 + \left(\frac{t}{\tau + r t}\right)^2.
\label{doft}
\ee
Here $d_0$ is the distance at the point of closest approach, $\tau$ is
a measure of the pulse duration, and $r$ is a dimensionless measure of
the time asymmetry of the pulse.  This form describes the generic
behavior of any pulse for times near the pulse maximum ($t = 0$).
Note that for large $|t|$, and for $r \ne 0$, the distance $d(t)$ will
saturate, and has a singularity for negative $t$.  We have taken $r$
to be small enough so that the dots decouple long before this leads to
any difficulty.

\end{multicols}

\begin{figure}[t]
\centerline{\psfig{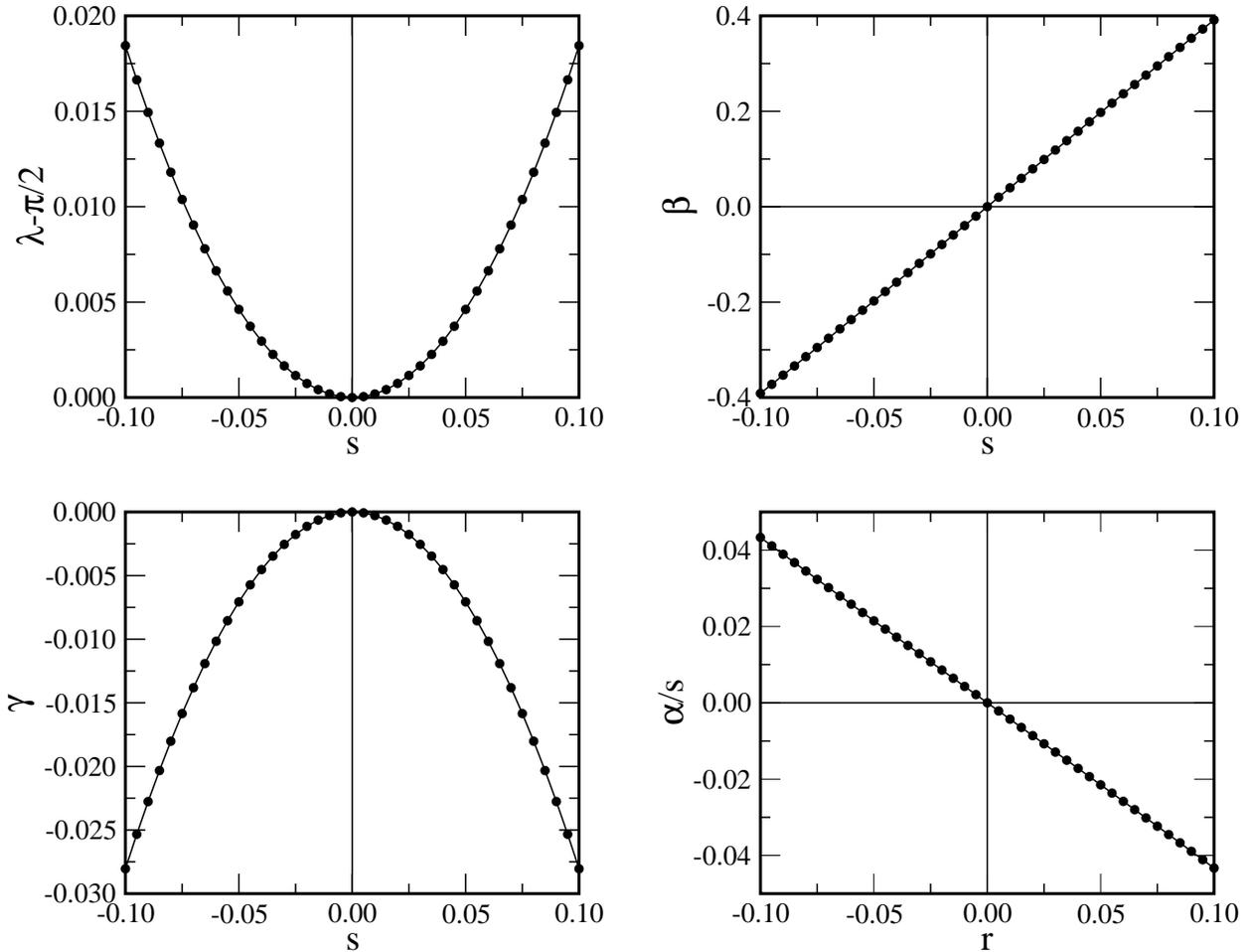}}\vskip.15in
\caption{Parameters appearing in the effective spin Hamiltonian
derived from pulses depicted in Fig.~\ref{dpulse}.  The parameters
$\alpha$, $\beta$ and $\gamma$ are shown as functions of $s$ for the
case $r=0$ (time-symmetric pulses).  For $\alpha$ the quantity
$\alpha/s$ is plotted vs. $r$.  We have verified that the ratio
$\alpha/s$ is essentially independent of $s$ for all values we have
considered ($|s| \le 0.1$).}
\label{heff}
\end{figure}

\begin{multicols}{2}

For our calculations, we work in zero magnetic field and take
$\hbar\omega_0 = 3$ meV and $d_0 = 1$, corresponding to $a \simeq 20$
nm at closest approach.  The resulting time dependences of the
parameters in $H_{HM}$ are shown in Fig.~\ref{dpulse}.  Note that the
spin-orbit matrix element plotted in this figure is $l_{SO}$, while
the spin-orbit matrix element appearing in $H_{HM}$ is ${\bf P} = s
l_{SO} \hat z $ where $s$ is the dimensionless measure of spin-orbit
coupling introduced in Sec.~II.

For a given pulse $H_{HM}(t)$ we integrate the time-dependent
Schr\"odinger equation to obtain the evolution operator $U$ for the
full pulse.  If the pulse is adiabatic then the matrix elements of $U$
which couple the singly occupied states $|S_1\rangle$ and
$|T_0\rangle$ to the doubly occupied state $|S_2\rangle$ can be made
negligibly small.\cite{schliemann} The quantum gate is then obtained
by simply truncating $U$ to the $4\times 4$ matrix acting on the
two-qubit Hilbert space.  By taking the $\log$ of this matrix we
obtain $\tau H = i \log U$ and thus the parameters $\lambda, \alpha,
\beta, \gamma$.  Note that when calculating $\log U$, there are branch
cuts associated with each eigenvalue of $U$, and as a consequence
$\tau H$ is not uniquely determined.  We resolve this ambiguity by
requiring that as the pulse height is reduced to zero and $U$ goes
continuously to the identity that $\tau H \rightarrow 0$ without
crossing any branch cuts.

We fix the pulse width $\tau$ by requiring that if we turn off
spin-orbit coupling ($s=0$) we obtain a $\lambda = \pi/2$ pulse,
i.e. a square-root of swap.  For the parameters used here we find this
corresponds to taking $\tau = 23.9/\omega_0 \simeq 5$ ps.  We have
checked that these pulses are well into the adiabatic regime.  The
magnitudes of the matrix elements coupling singly occupied states to
the doubly occupied state $|S_2\rangle$ are on the order of $|\langle
S_1 | U | S_2\rangle| \sim 10^{-6}$ and $|\langle T_0 | U |
S_2\rangle| \sim s 10^{-6}$.

\vskip 0.1in

\begin{table}[h]
\vskip 0.3in
\caption{Symmetry properties of the pulse parameters $r$ and $s$, and
gate parameters $\lambda$, $\alpha$, $\beta$ and $\gamma$ under parity
$P$ and time reversal $T$.}
\begin{tabular}{ccc|ccc|ccccc}
&  & & $r$  & $s$ & & $\lambda$ & $\alpha$ & $\beta$ & $\gamma$ \\  
\tableline
& P &  & $+$  & $-$ & & $+$  & $-$  &  $-$  &  $+$ \\ 
& T &  & $-$  & $+$ & & $+$  & $-$  &  $+$  &  $+$
\end{tabular}
\end{table}

Once $\tau$ is fixed, there are two parameters characterizing each
pulse, $s$ and $r$, and four parameters characterizing the resulting
gate, $\lambda$, $\alpha$, $\beta$, and $\gamma$.  The transformation
properties of these parameters under parity ($P$) and time reversal
($T$) are summarized in Table I.  These properties follow from the
fact that (i) under time reversal ${\bf S}_\mu \rightarrow - {\bf
S}_\mu$ and $r \rightarrow -r$, while ${\bf P} = s l_{SO}{\hat {\bf
z}}$ is invariant, and (ii) under parity ${\bf S}_A \leftrightarrow
{\bf S}_B$ and ${\bf P} \rightarrow -{\bf P}$, while $r$ is invariant.
Note that, as defined in Sec.~II, the parameter $s$ is positive. Here
we allow $s$ to change sign when the direction of the vector ${\bf P}$
is reversed, thus under parity $s \rightarrow -s$.

These symmetry properties imply that if $s$ and $r$ are small, the
parameters of the effective Hamiltonian will be given approximately by
\be
\alpha &\simeq& C_\alpha r s, \\
\beta &\simeq& C_\beta s, \\
\gamma &\simeq& C_\gamma s^2, \\
\lambda &\simeq& \lambda_0 + C_\lambda s^2,
\ee
where the coefficients should be of order 1. For the pulses we
consider here $\lambda_0 = \pi/2$.

The results of our calculations are shown in Fig.~\ref{heff}.  Each
point corresponds to a separate numerical run.  The plots for
$\lambda$, $\beta$ and $\gamma$ show their dependence on $s$ when
$r=0$.  The dependence of the parameter $\alpha$ on pulse asymmetry is
shown by plotting $\alpha/s$ vs. $r$.  For the $s$ values we have
studied, up to $|s| = 0.1$, the numerical results for $\alpha/s$ are
essentially independent of $s$ for a given $r$.  These results are
clearly consistent with the above symmetry analysis.

Now consider carrying out a CNOT gate using the scheme reviewed in
Sec.~III.  For this construction to work it is necessary that $\lambda
= \pi/2$. In our calculations we have fixed $\tau$ so that $\lambda =
\pi/2$ for $s=r=0$.  Thus, when spin-orbit coupling is included
\be
\lambda \simeq \pi/2 + C_\lambda s^2.
\ee
In order to keep $\lambda = \pi/2$ it will therefore be necessary to
adjust the pulse width $\tau$ slightly to correct for spin-orbit
effects.  

The central result of this paper is summarized by the equation
\be
\alpha \simeq C_\alpha r s.
\label{alpha}
\ee
As shown in Sec.~III, any nonzero $\alpha$ will lead to corrections to
the CNOT construction. For time-symmetric pulses $r=0$ and these
corrections will vanish.  Equation (\ref{alpha}) can then be used to
estimate the errors due to any time asymmetry of the pulse, and to put
design restrictions on the allowed tolerance for such asymmetry.

It is important to note that while the results presented here are for
a specific model, all of the key arguments are based on symmetry and
so are quite general.  Given any time-reversal invariant two-qubit
system with axial symmetry, if pulsed adiabatically in a
time-symmetric way the resulting gate will have the form
(\ref{axial}).

If the pulse is not axially symmetric, e.g. if the ratio $f_D/f_R$
varies during the pulse, then time-symmetric pulsing will still
restrict the resulting gate to be invariant under time reversal.
Thus, up to an irrelevant overall phase, this gate will necessarily
have the form
\be
U = \exp -i \lambda ({\bf S}_A \cdot {\bf S}_B
+\mbox{\boldmath{$\beta$}} \cdot {\bf S}_A \times {\bf S}_B
+ {\bf S}_A \cdot {\bf I}\!{\mbox{\boldmath{$\Gamma$}}} \cdot {\bf S}_B).
\ee
Here ${\bf I}\!{\mbox{\boldmath{$\Gamma$}}}$ is a symmetric tensor
which will, in general, deviate from the axial form of the $\gamma$
term in (\ref{axial}) leading to corrections to the CNOT construction.
However, because ${\bf I}\!{\mbox{\boldmath{$\Gamma$}}}$ is even under
parity it will still be second order in spin-orbit
coupling,\cite{bonesteel01} and thus the deviations from (\ref{axial})
will also be second order.  We conclude that even in the absence of
axial symmetry, the corrections to the CNOT construction will be
second order in spin-orbit coupling, rather than first order.

\section{Conclusions}

In this paper we have studied spin-orbit corrections to exchange-based
quantum gates, emphasizing symmetry arguments.  In particular, we have
shown that adiabatic time-symmetric pulsing of any Hamiltonian which
(i) describes two well defined spin-1/2 qubits at the beginning and
end of the pulse, (ii) is time-reversal symmetric at all times during
the pulse, and (iii) is axially symmetric in spin space with a fixed
symmetry axis, will automatically produce a gate of the form
(\ref{axial}).  Together with single qubit rotations, for $\lambda =
\pi/2$ this gate can then be used in a simple CNOT construction. This
result is quite general.

As a specific example we have studied a GaAs double quantum dot system
within the Hund-Mulliken approximation.  In this approximation
spin-orbit coupling enters as a small spin precession when an electron
tunnels between dots.  If the direction of this precession axis is
constant throughout the pulse the resulting gate will be axially
symmetric and have the form (\ref{axialnt}).  The deviation of this
gate from the desired gate (\ref{axial}) is then characterized by a
single dimensionless parameter $\alpha$ which spoils the CNOT
construction.  Using symmetry arguments, as well as numerical
calculations, we have shown that $\alpha \simeq C_\alpha s r$ where $s$ and
$r$ are, respectively, dimensionless measures of spin-orbit coupling
and time asymmetry of the pulse.  Thus time-symmetric pulsing ($r=0$)
ensures the anisotropic corrections will have the desired form.

In any system without spatial inversion symmetry, spin-orbit coupling
will inevitably lead to anisotropic corrections to the exchange
interaction between spins.  According to current
estimates,\cite{aharonov} fault-tolerant quantum computation will
require realizing quantum gates with an accuracy of one part in
$10^4$.  Thus, even if spin-orbit coupling is weak, the design of any
future quantum computer which uses the exchange interaction will have
to take these anisotropic corrections into account.  We believe the
symmetry based analysis presented in this paper provides a useful
framework for studying these effects.

\acknowledgements DS and NEB acknowledge support from the National
Science Foundation through NIRT Grant No.\ DMR-0103034.  DPDV is
supported in part by the National Security Agency and the Advanced
Research and Development Activity through Army Research Office
contract number DAAD19-01-C-0056.  He thanks the Institute for Quantum
Information at Cal Tech (supported by the National Science Foundation
under Grant. No. EIA-0086038) for its hospitality during the initial
stages of this work.  DL thanks Swiss NSF, NCCR Nanoscience, DARPA and
ARO.

\end{multicols}


\begin{references}

\bibitem{loss} D. Loss and D.P. DiVincenzo, Phys. Rev. A {\bf 57},
120 (1998).

\bibitem{burkard99_1} G. Burkard, D. Loss, and D.P. DiVincenzo,
Phys. Rev. B {\bf 59}, 2070 (1999).

\bibitem{burkard99_2} G. Burkard, D. Loss, D.P. DiVincenzo, and
J.A. Smolin, Phys. Rev. B {\bf 60}, 11404 (1999).

\bibitem{hu} X. Hu and S. Das Sarma, Phys. Rev. A {\bf 61}, 062301
(2000).

\bibitem{schliemann} J. Schliemann, D. Loss and A.H. MacDonald,
Phys. Rev. B {\bf 63}, 085311 (2001).

\bibitem{bacon} D. Bacon, J. Kempe, D.A. Lidar, and K.B. Whaley,
Phys. Rev. Lett. {\bf 85}, 1758 (2000).

\bibitem{divincenzo} D.P. DiVincenzo, D. Bacon, J. Kempe, G. Burkard
and K.B. Whaley, Nature {\bf 408}, 339 (2000).

\bibitem{burkard01} G. Burkard and D. Loss,  Phys. Rev. Lett. {\bf 88}, 047903
(2002).

\bibitem{wu} L.-A. Wu and D. Lidar, Phys. Rev. A {\bf 66}, 062314
(2002).

\bibitem{kavokin} K.V. Kavokin, Phys. Rev. B {\bf 64} 075305 (2001);
cond-mat/0212347.

\bibitem{gorkov} L.P. Gor'kov and P.L. Krotkov, Phys. Rev. B {\bf 67},
033203 (2003).

\bibitem{bonesteel01} N.E. Bonesteel, D. Stepanenko, and
D.P. DiVincenzo, Phys. Rev. Lett. {\bf 87}, 207901 (2001).

\bibitem{dresselhaus} G. Dresselhaus, Phys. Rev. {\bf 100}, 580
(1955).

\bibitem{dyakonov} M. I. Dyakonov and V. Y. Kachorovskii, Sov. Phys.
Semicond. {\bf 20}, 110 (1986).

\bibitem{rashba} E. L. Rashba, Fiz. Tv. Tela (Leningrad) {\bf 2}, 
1224 (1960) [Sov. Phys. Solid State {\bf 2}, 1109 (1960)]; Y.A. Bychkov
and E.I. Rashba, J. Phys. C {\bf 17}, 6039 (1984).

\bibitem{schliemann03} J. Schliemann, J.C. Egues, and D. Loss,
Phys. Rev. Lett. {\bf 90}, 146801 (2003).

\bibitem{gottfried} For an excellent discussion of time-reversal
symmetry see, ``Quantum Mechanics'', K. Gottfried, Addison-Wesley,
1989, pp. 314-322.

\bibitem{aharonov} D. Aharonov and M. Ben-Or, quant-ph/9906129.

\end{references}
\end{document}